\documentclass[iop]{emulateapj}

\usepackage{graphicx}
\def\dlambda{$\lambda\lambda$}
\def\kms{km~s$^{-1}$} 

\slugcomment{Accepted to ApJL on May 2, 2013}

\shorttitle{SN\,2012au: The Golden Link}
\shortauthors{Milisavljevic et al.}

\begin{document}

\def\cfa{1}
\def\carn{2}
\def\dartmouth{3}
\def\lco{4}
\def\cu{5}
\def\osu{6}
\def\arh{7}

\title{SN 2012au: A Golden Link Between Superluminous Supernovae\\and
  Their Lower-Luminosity Counterparts}

\author{Dan~Milisavljevic\altaffilmark{\cfa}, 
Alicia~M.~Soderberg\altaffilmark{\cfa},
Raffaella~Margutti\altaffilmark{\cfa},
Maria~R.~Drout\altaffilmark{\cfa},
G.~Howie~Marion\altaffilmark{\cfa},
Nathan~E.~Sanders\altaffilmark{\cfa},
Eric~Y.~Hsiao\altaffilmark{\carn},
Ragnhild~Lunnan\altaffilmark{\cfa},
Ryan~Chornock\altaffilmark{\cfa},
Robert~A.~Fesen\altaffilmark{\dartmouth},
Jerod~T.~Parrent\altaffilmark{\dartmouth,\lco},
Emily~M.~Levesque\altaffilmark{\cu},
Edo~Berger\altaffilmark{\cfa},
Ryan~J.~Foley\altaffilmark{\cfa},
Pete~Challis\altaffilmark{\cfa},
Robert~P.~Kirshner\altaffilmark{\cfa},
Jason~Dittmann\altaffilmark{\cfa},
Allyson~Bieryla\altaffilmark{\cfa},
Atish~Kamble\altaffilmark{\cfa},
Sayan Chakraborti\altaffilmark{\cfa},
Gisella~De~Rosa\altaffilmark{\osu},
Michael~Fausnaugh\altaffilmark{\osu},
Kevin~N.~Hainline\altaffilmark{\dartmouth},
Chien-Ting~Chen\altaffilmark{\dartmouth},
Ryan~C.~Hickox\altaffilmark{\dartmouth},
Nidia~Morrell\altaffilmark{\carn},
Mark~M.~Phillips\altaffilmark{\carn},
Maximilian~Stritzinger\altaffilmark{\arh}
} 

\altaffiltext{\cfa}{Harvard-Smithsonian Center for Astrophysics, 60
  Garden Street, Cambridge, MA, 02138. \\ Electronic address:
  dmilisav@cfa.harvard.edu} 
\altaffiltext{\carn}{Carnegie Observatories,
  Las Campanas Observatory, Colina El Pino, Casilla 601, Chile}
\altaffiltext{\dartmouth}{6127 Wilder Lab, Department of Physics \& Astronomy, Dartmouth
                 College, Hanover, NH, 03755}
\altaffiltext{\lco}{Las Cumbres Observatory Global Telescope Network, Goleta, CA, US}

\altaffiltext{\cu}{CASA, Department of Astrophysical and Planetary
        Sciences, University of Colorado, 389-UCB, Boulder, CO 80309, USA}
\altaffiltext{\osu}{Department of Astronomy, The Ohio State
  University, 140 West 18th Avenue, Columbus, OH 43210, USA}
\altaffiltext{\arh}{Department of Physics and Astronomy, Aarhus
        University, Ny Munkegade 120, DK-8000 Aarhus C, Denmark}

\begin{abstract}

  We present optical and near-infrared observations of SN\,2012au, a
  slow-evolving supernova (SN) with properties that suggest a link
  between subsets of energetic and H-poor SNe and superluminous
  SNe. SN\,2012au exhibited conspicuous SN Ib-like \ion{He}{1} lines
  and other absorption features at velocities reaching $\approx
  2\times10^4$ \kms\ in its early spectra, and a broad light curve
  that peaked at $M_B = -18.1$\,mag. Models of these data indicate a
  large explosion kinetic energy of $\sim 10^{52}$\,erg and $^{56}$Ni
  mass ejection of $M_{\rm Ni} \approx 0.3 M_{\odot}$ on par with
  SN\,1998bw. SN\,2012au's spectra almost one year after explosion
  show a blend of persistent \ion{Fe}{2} P-Cyg absorptions and nebular
  emissions originating from two distinct velocity regions. These
  late-time emissions include strong [\ion{Fe}{2}], [\ion{Ca}{2}],
  [\ion{O}{1}], \ion{Mg}{1}], and \ion{Na}{1} lines at velocities $\ga
  4500$ \kms, as well as \ion{O}{1} and \ion{Mg}{1} lines at
  noticeably smaller velocities $\la 2000$ \kms. Many of the late-time
  properties of SN\,2012au are similar to the slow-evolving hypernovae
  SN\,1997dq and SN\,1997ef, and the superluminous SN\,2007bi. Our
  observations suggest that a single explosion mechanism may unify all
  of these events that span $-21 \la M_B \la -17$~mag. The aspherical
  and possibly jetted explosion was most likely initiated by the core
  collapse of a massive progenitor star and created substantial
  high-density, low-velocity Ni-rich material.

\end{abstract}


\section{Introduction}
\label{sec:Intro}

Recent transient surveys (e.g., the Panoramic Survey Telescope and
Rapid Response System, the Palomar Transient Factory, the Catalina
Real-Time Transient Survey [CRTS], the Texas Supernova Search
now operating as the ROTSE Supernova Verification Project) and growing
support from amateur observers have uncovered ever-increasing
diversity in the observational properties of supernovae (SNe). Indeed,
the standard classification system Type I and Type II originally
proposed by \citet{Minkowski41} has branched considerably from its
binary roots, and is continually being updated in the face of new
objects that bridge subtypes and extend luminosity ranges.

Some of the more luminous H- and He-poor SNe discovered by these
efforts have garnered especial attention because of their connection
with long-duration gamma-ray bursts (GRBs; see \citealt{Hjorth12} for
a recent review). Strong evidence for this relationship came from
multi-wavelength observations of SN\,1998bw, a broad-lined SN Ic
coincident with GRB980425, and belonging to an energetic class of
explosions reaching $\sim 10^{52}$~erg called hypernovae
\citep{Galama98,Iwamoto98}. The handful of energetic broad-lined SN Ic
that have been well-observed vary considerably in their properties and
are not always accompanied with GRBs, e.g., SN\,1997ef
\citep{Iwamoto00}.

superluminous SNe (SLSNe) with absolute magnitudes of $\la -21$ are
recent members of the growing SN classification zoo
\citep{Quimby11}. H-poor examples can be even more powerful than SNe
Ic \citep{Gal-Yam12}, and there has been considerable effort to
understand their nature and relationship with more typical SNe.  One
of the first examples of the H-poor variety was SN\,2007bi
\citep{Gal-Yam09,Young10}, which was originally suggested to be the
result of the pair-instability explosion mechanism
\citep{Barkat67}. Alternative interpretations, however, have been offered
\citep{Dessart12,Dessart13}, including those involving a
central engine magnetar \citep{Kasen10}, interaction with
circumstellar shells \citep{Chatzopoulos12}, or Fe core collapse
\citep{Moriya10,Yoshida11}. To date, no SLSN has been observationally
connected to a GRB.

\begin{figure*}[htp!]
\centering 

\includegraphics[width=\linewidth]{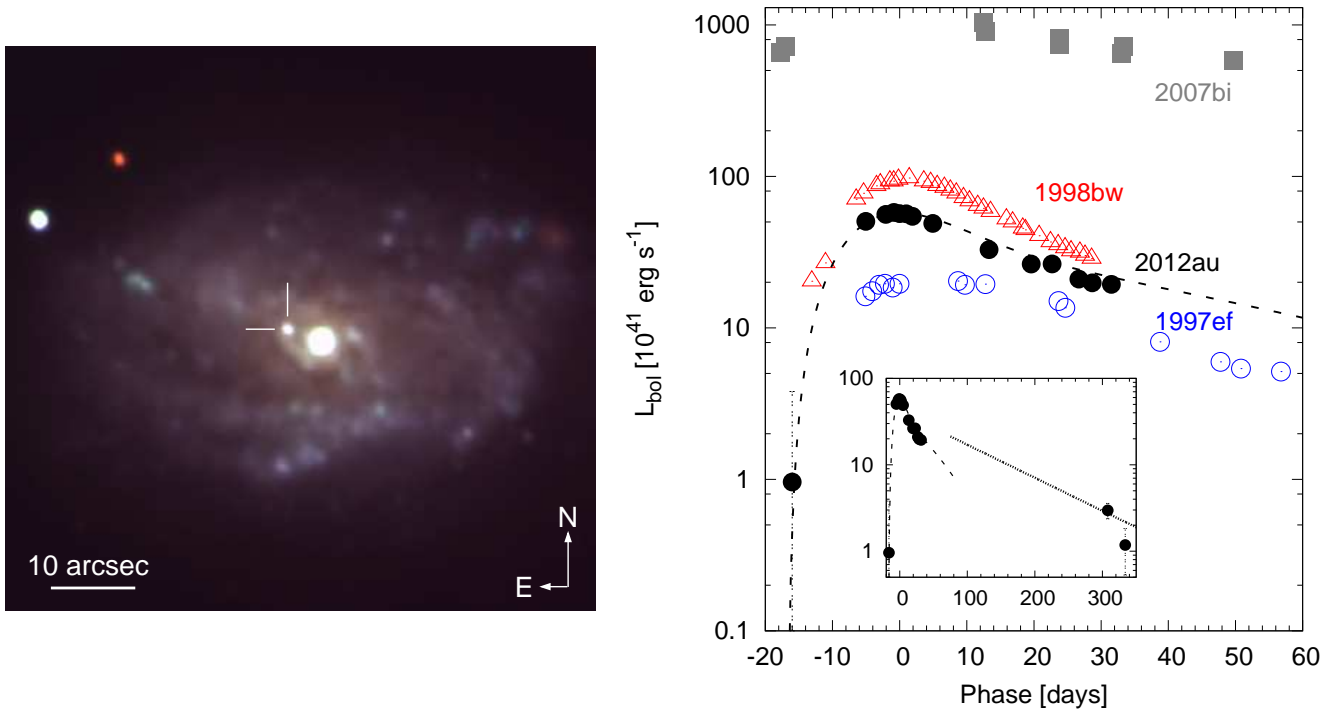}

\caption{Left: Image of the region around SN\,2012au (marked) and its
  host galaxy NGC\,4790 made from MMTcam $g'r'i'$ images obtained 2013
  February 18. Right: Reconstructed pseudo-bolometric light curve of
  SN\,2012au and our radioactive decay diffusion model fit for
  photospheric (dashed line) and nebular (dotted line, inset)
  epochs. Phase is in the rest-frame and with respect to maximum
  light. Also shown are bolometric light curves of SN\,1997ef
  \citep{Nomoto00}, SN\,1998bw \citep{Maeda06}, and SN\,2007bi
  \citep{Gal-Yam12}.}

\label{fig:lcurves}
\end{figure*}

Here we report photometric and spectroscopic data on SN\,2012au, an
object that suggests a link between some subtypes of energetic and
H-poor SNe and SLSNe. Although initial spectroscopic observations
showed prominent helium absorption features of an otherwise ordinary
SN~Ib \citep{Silverman12,Soderberg12}, continued monitoring of
SN\,2012au through to nebular stages ($t > 250$ d) revealed
extraordinary emission properties in its optical and near-infrared
spectra that prompted this {\it Letter}.  A complementary analysis of
SN\,2012au's radio and X-ray emissions is forthcoming in Kamble et
al.\ (in prep).

%%%%%%%%%%%%%%%%%%%%%%%%%%%%%%%%%%%%%%%%%%%%%%%%%%%%%%%%%%%
\section{Observations and Results}
\label{sec:Observations}
%%%%%%%%%%%%%%%%%%%%%%%%%%%%%%%%%%%%%%%%%%%%%%%%%%%%%%%%%%%

\subsection{UV and Optical Photometry}

SN\,2012au was discovered on 2012 March 14 UT by the CRTS SNHunt
project \citep{Howerton12}. Figure~\ref{fig:lcurves} (left) shows the
region around the SN and its host galaxy NGC\,4790 ($D = 23.5 \pm 0.5$
Mpc; \citealt{Theureau07}). SN\,2012au is located at coordinates
$\alpha = 12^{\rm h}54^{\rm m}52\fs18$ and $\delta =
-10\degr14'50\farcs2$ (J2000.0), which is is less than 600 pc away in
projection from the center of NGC\,4790's bright central nucleus.

Ultra-violet (UV) and optical observations with the \emph{Swift}-UVOT
instrument began 2012 March 15 and continued through to 2012 April 21
(PI P.\ Brown).  Data were acquired using six broad band filters and
have been analyzed following standard procedures.  The details of
these \emph{Swift}-UVOT observations are provided in
Table~\ref{tab:photometry}.

Two epochs of late-time $r'$-band photometry of SN\,2012au were
obtained. A sequence of $3 \times 300$\,s dithered images were taken
on 2013 January 23 using the 2.4m Hiltner telescope at MDM Observatory
with the OSMOS
instrument\footnote{http://www.astronomy.ohio-state.edu/$\sim$martini/osmos/}
and MDM4k detector, and a sequence of $3 \times 120$\,s dithered
images were taken on 2013 Feb 18 using the 6.5m MMT with the MMTcam
instrument\footnote{http://www.cfa.harvard.edu/mmti/wfs.html}. Images
were bias-subtracted, flat-fielded, and stacked following standard
procedures using the IRAF\footnote{IRAF is distributed by the National
Optical Astronomy Observatory, which is operated by the Association of
Universities for Research in Astronomy (AURA) under cooperative
agreement with the National Science Foundation.} software.

Emission from SN\,2012au dropped $\la 4$~mag between maximum light and
our first late-time measurement 308 days later. This decline is
considerably slower than other SNe Ib/c; e.g., SN\,1998bw declined by
$\approx 6.5$~mag over the same time period \citep{Patat01}. The
longevity of the brightness made it impossible to obtain subtraction
templates that could completely remove contaminating emission from the
underlying host galaxy. Photometric measurements were thus made using
the \texttt{sextractor} software \citep{Bertin96} and checked manually
using point-spread-function fitting techniques.

\subsection{A Model of the Bolometric Light Curve}

We constructed a pseudo-bolometric light curve using the
\emph{Swift}-UVOT photometry. Observations were corrected assuming
$R_V = A_V/E(B-V) =3.1$ and a total reddening of $E(B-V) = 0.063$
mag. This estimate combines reddening due to the Milky Way,
$E(B-V)_{mw} = 0.043$ mag \citep{Schlafly11}, and host internal
extinction of $E(B-V)_{host} = 0.02 \pm 0.01$ mag determined from
measuring the equivalent width (EW) of \ion{Na}{1}\,D absorption in
our optical spectra (Section \ref{sec:spectra}) and using the
relationship between EW and $E(B-V)$ described in \citet{Poznanski12}.

\begin{figure}[htp!]
\centering
\includegraphics[width=\linewidth]{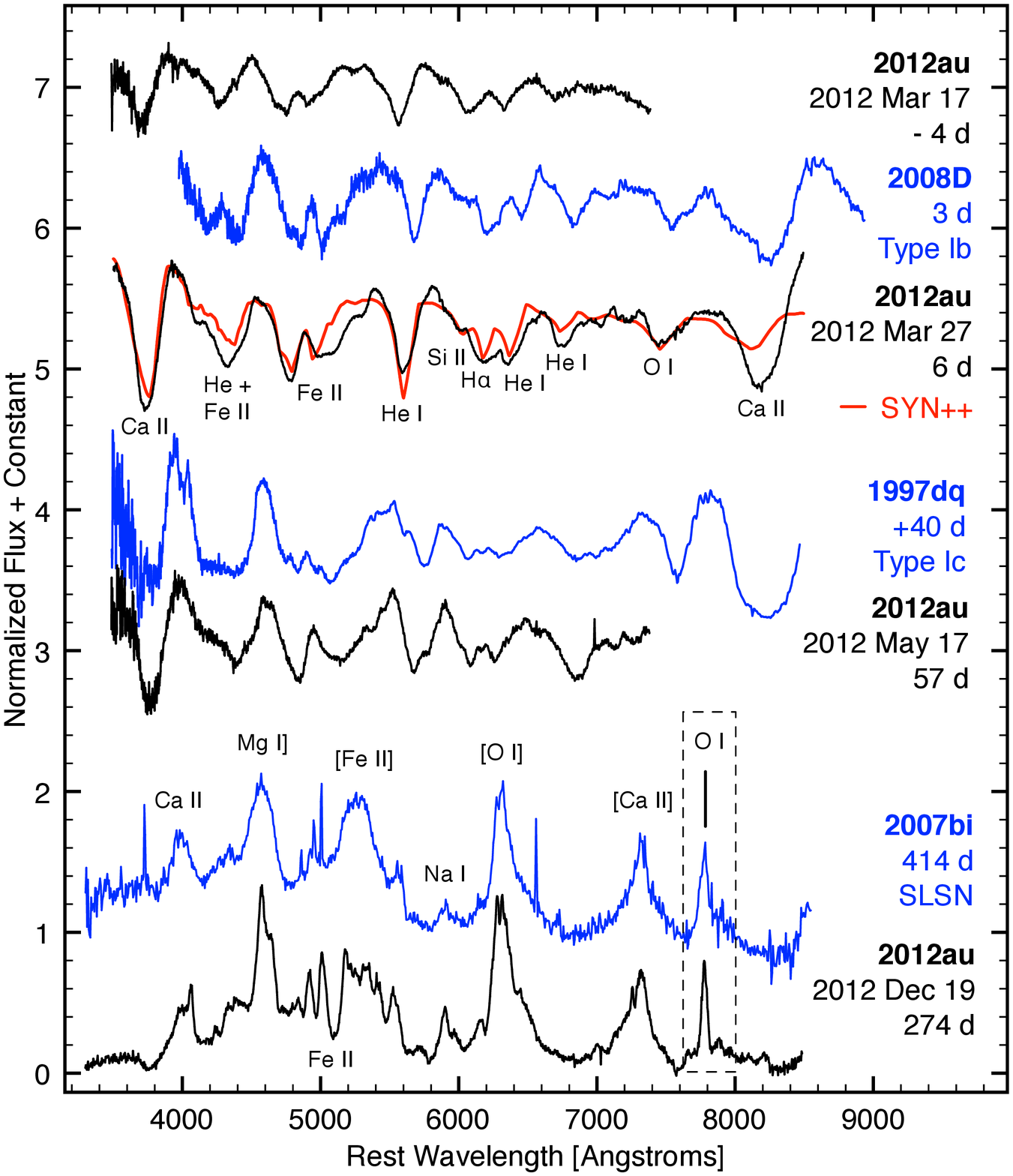}\\
\includegraphics[width=\linewidth]{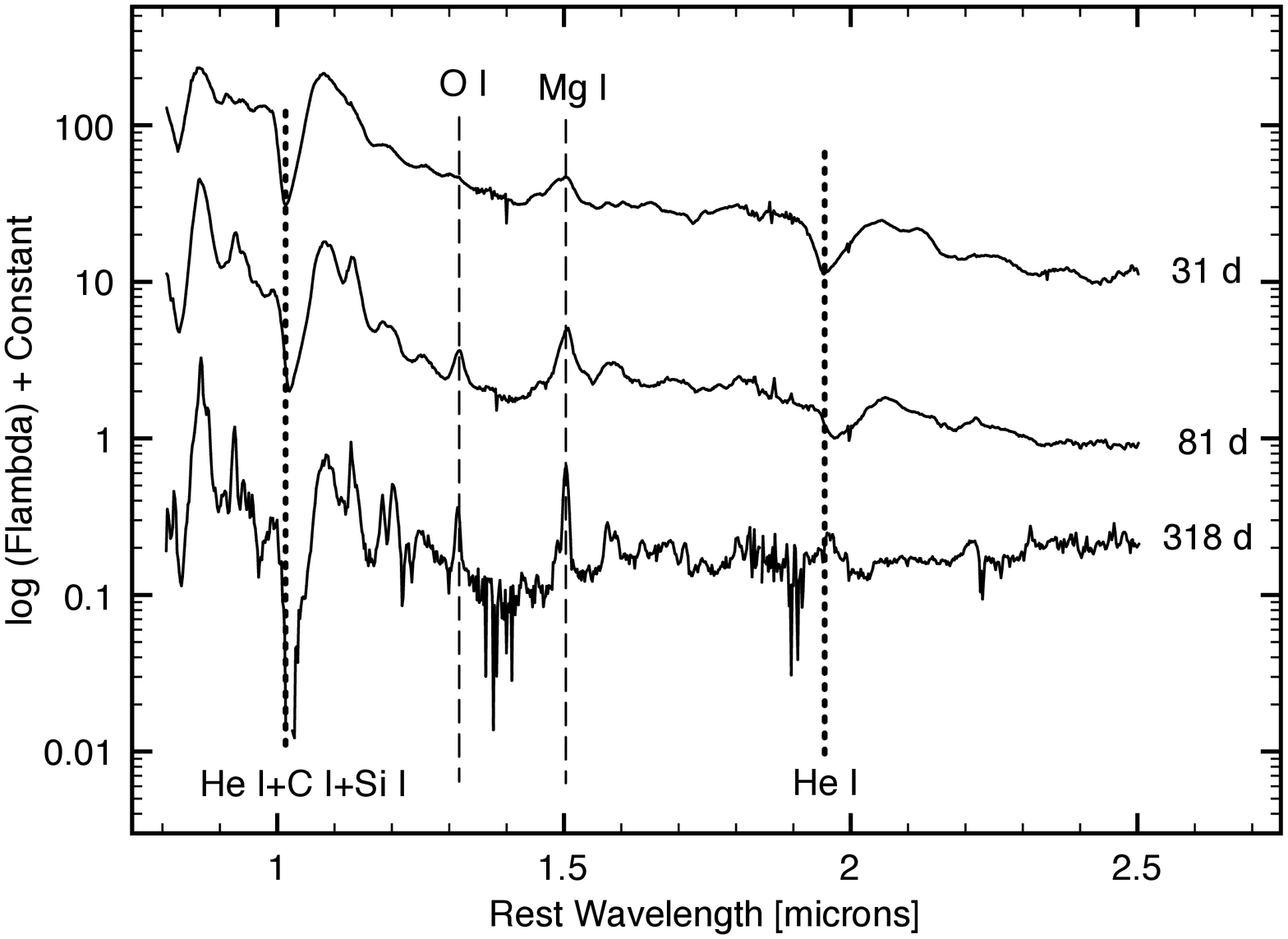}

\caption{Top: Optical spectra of SN\,2012au. In red is the SYN++
  synthetic spectrum. Also shown are SN\,2008D \citep{Soderberg08},
  SN\,1997dq \citep{Matheson01}, and SN\,2007bi \citep{Gal-Yam09}.
  Spectra have been normalized according to the procedure outlined in
  \citet{Jeffery07}. Bottom: Near-infrared spectra of SN\,2012au. The
  heavy dotted lines highlight \ion{He}{1} absorptions, and the light
  dashed lines highlight \ion{O}{1} and \ion{Mg}{1} emission
  features.}

\label{fig:spectra}
\end{figure}

The pseudo-bolometric light curve incorporates two values from
\citet{Howerton12}, one of which is a detection of the SN on 2012
March 4 that constrains the rise time from explosion to peak
brightness to be $\gtrsim$ 16 days. We summed the flux emitted in the
$uvw2$ through $v$ bands by means of a trapezoidal interpolation and
estimated the missing flux from the $RI$ pass-bands and the
near-infrared around the time of maximum light, $t_{\rm max}$.  Using
values observed in other SNe Ib/c (see \citealt{Valenti08a}), we
assumed that for $-5 < \Delta t_{\rm max} < 30$\,d, the $RI-$band flux
contribution varies from 25\% to 40\% and the near-infrared band
contribution varies from 15\% to 45\%.

The pseudo-bolometric light curve was modeled using assumptions in
\citet{Arnett82} and following procedures in \citet{Valenti08a} to
determine the SN's total nickel mass, $M_{\rm Ni}$, explosion kinetic
energy, $E_{\rm K}$, and total ejecta mass $M_{\rm ej}$. A constant
optical opacity k$_{opt}$ of $0.05$ cm$^{2}$ g$^{-1}$ was adopted (see
\citealt{Drout11} for details). Our best-fit model shown in
Figure~\ref{fig:lcurves} (right) uses the explosion date 2012 March
3.5, and a rise time of $16.5 \pm 1.0$~d. The peak bolometric
luminosity is $\sim 6 \times 10^{42}$ erg\,s$^{-1}$, and the estimates
of the explosion parameters are $M_{\rm Ni} \approx 0.3 M_{\odot}$,
$M_{\rm ej} \approx 3 - 5 M_{\odot}$, and $E_{\rm K} \sim 1 \times
10^{52}$ erg.

We break the degeneracy between $E_{\rm K}$ and $M_{\rm ej}$ using an
estimate of the ejecta velocity measured from our optical spectra
around $t_{\rm max}$ to be $\approx 1.5 \times 10^4$ km s$^{-1}$
(Section~\ref{sec:spectra}).  The value of $M_{\rm Ni}$ is consistent
with independent estimates derived from the late-time photometry under
the assumption of full gamma-ray trapping (Figure~\ref{fig:lcurves},
right, inset).  These explosion parameters are comparable to those
estimated for the hypernovae SN\,1998bw ($M_{\rm Ni} \sim
0.4\;M_{\odot}$ and $E_{\rm K} \sim 2 \times 10^{52}$\,erg;
\citealt{Maeda06}) and SN\,1997ef ($M_{\rm Ni} \sim 0.16\;M_{\odot}$
and $E_{\rm K} \sim 1 \times 10^{52}$\,erg; \citealt{Iwamoto00}).

We note that SN\,2012au may exhibit properties that are not strictly
incorporated in the \citet{Arnett82} model we have used. Opacities as
high as $0.2$ cm$^{2}$ g$^{-1}$ could apply that increase the $M_{\rm
Ni}$/$M_{\rm ej}$ ratio to values that run contrary to model
assumptions, and optical spectra (Section~\ref{sec:late-time}) suggest
possible deviations from the model's spherically symmetric
conditions. Future work may look to apply more sophisticated models to
address these additional factors.

\subsection{Optical Spectroscopy ($t < 60$ day)}
\label{sec:spectra}

Low-resolution optical spectra of SN\,2012au were obtained from March
2012 through Feb 2013. Many of these observations were made with the
F.~ L.\ Whipple Observatory 1.5-m telescope mounted with the FAST
instrument \citep{Fabricant98}. Additional observations were obtained
with the MMT 6.5m telescope using the Blue Channel instrument
\citep{Schmidt89}, and the 2.4m MDM telescope using the OSMOS
instrument.

Standard procedures using the IRAF software were followed and
flux calibrations were made using our own IDL routines. A recession
velocity of 1295 \kms\ determined from overlapping nebular H$\alpha$
emission has been removed from all spectra. Line identifications and
estimates of expansion velocities of the photospheric spectra were
made with the supernova spectrum synthesis code SYN++
\citep{Thomas11}.

Four epochs of our optical spectra of SN\,2012au are plotted in
Figure~\ref{fig:spectra} (top). The reported phase is with respect to
$v$-band maximum on 2012 March 21. Our earliest spectrum obtained 2012
Mar 17 ($\Delta t_{\rm max} = -4$\,d) shows features associated with
\ion{He}{1}, \ion{Fe}{2}, \ion{Si}{2}, \ion{Ca}{2}, \ion{Na}{1}, and
\ion{O}{1} exhibiting velocities of $(1.8-2.0) \times
10^4$~\kms. There is weak evidence of H$\alpha$ absorption.

We show a slightly later spectrum obtained 2012 Mar 27 ($\Delta t_{\rm
max} = +6$\,d) in Figure~\ref{fig:spectra} (top) illustrating how the
identified ions are generally consistent with a Type Ib classification
(e.g., SN\,2008D) though the velocities in SN\,2012au are relatively
high. Approximately two months later on 2012 Mar 17 ($\Delta t_{max} =
+57$\,d), the same ions are detected but the Type Ib classification is
less appropriate. The observed velocities ($\sim 7000$ \kms) remain
above average relative to other SNe Ib/c at this epoch, the P-Cyg
absorptions broaden, and the spectrum resembles those of SN\,1997dq
and SN\,1997ef \citep{Matheson01}.

\begin{figure}[htp!]
\centering
\includegraphics[width=0.544\linewidth]{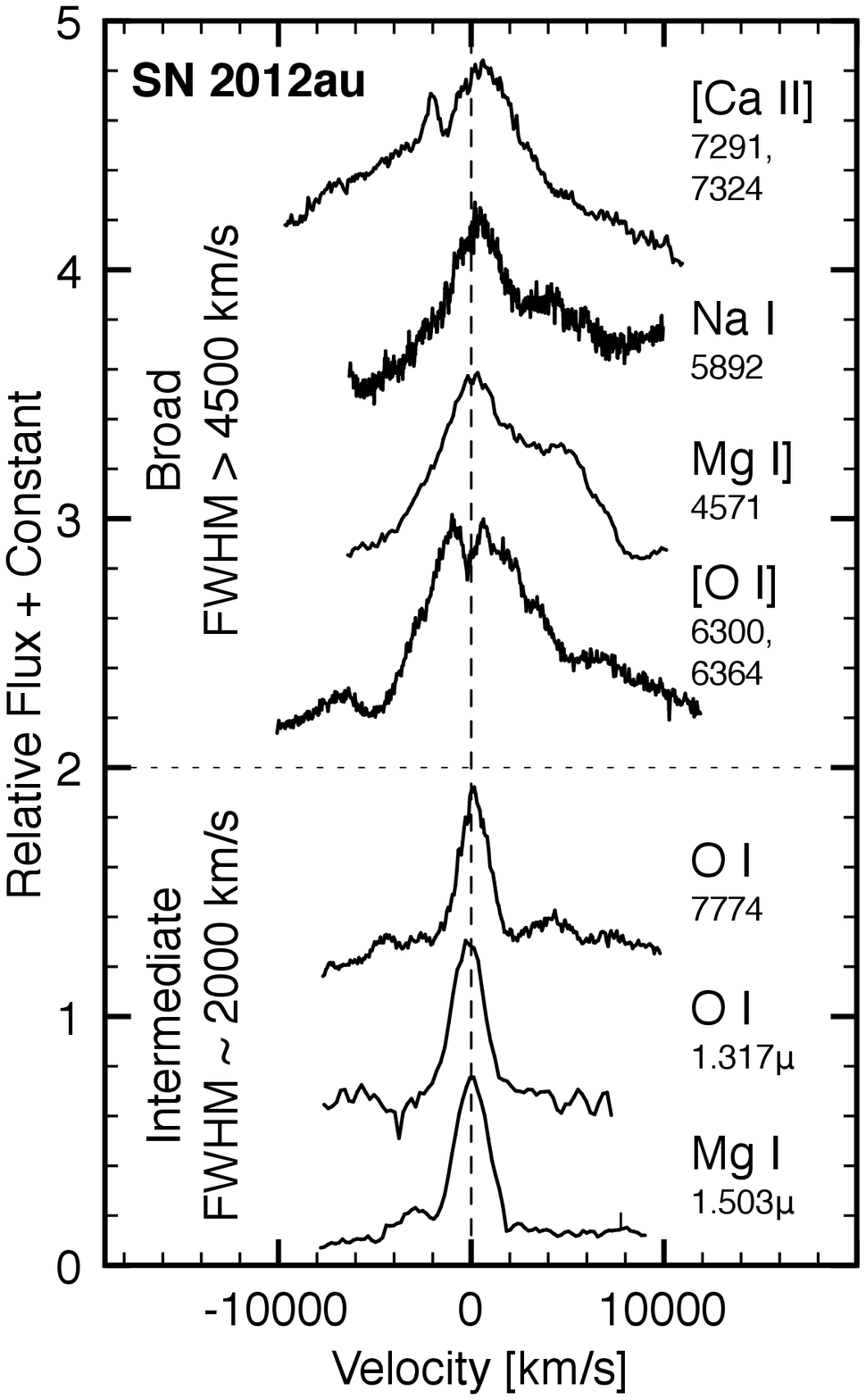}
\includegraphics[width=0.428\linewidth]{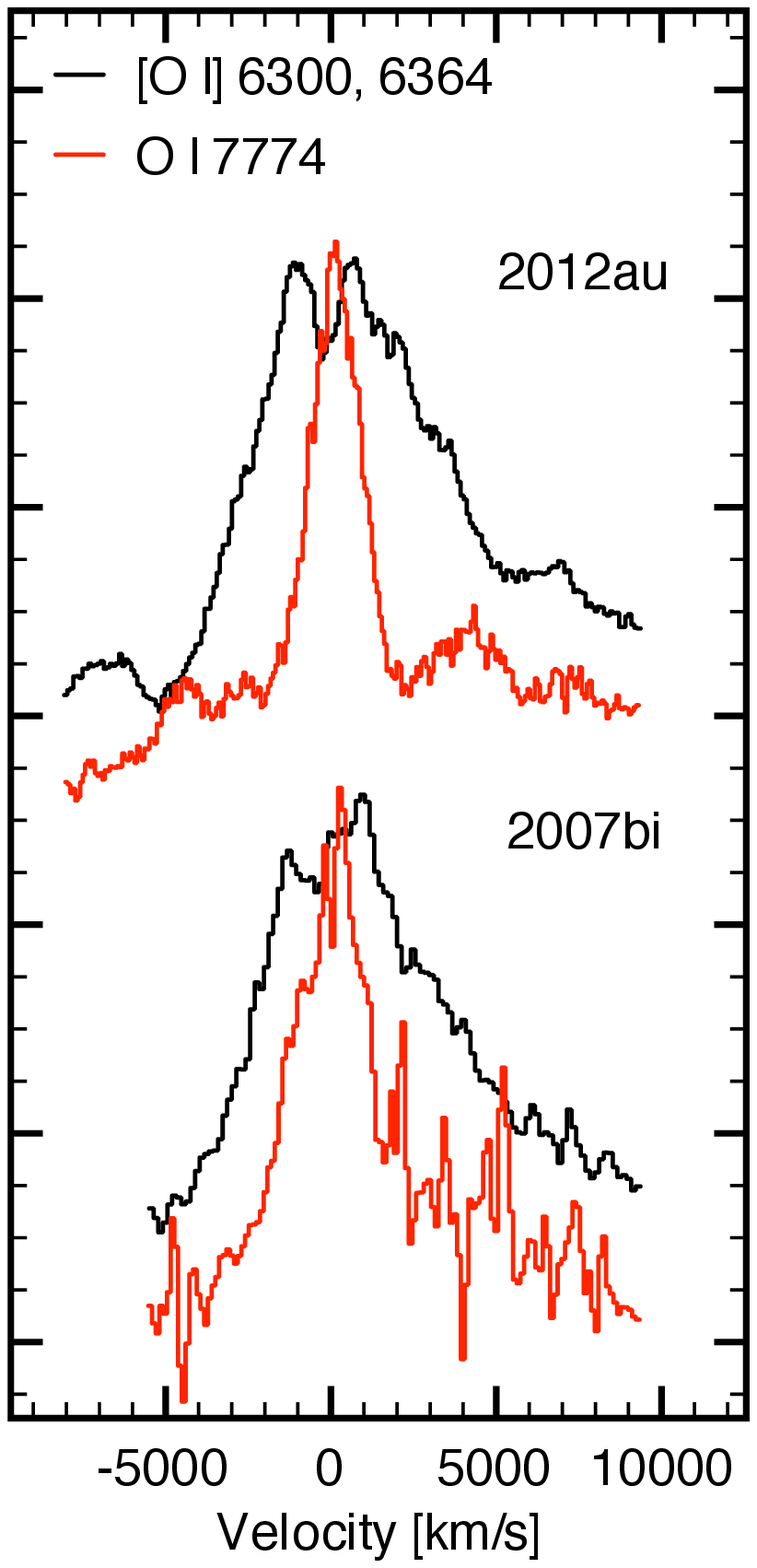}

\caption{Left: Late-time emission line profiles of
  SN\,2012au. Profiles are grouped by broad (FWHM $\ga 4500$ \kms) and
  intermediate (FWHM $\sim 2000$ \kms) widths. Right: The \ion{O}{1}
  $\lambda$7774 and [\ion{O}{1}] \dlambda 6300, 6364 emission line
  profiles of SN\,2012au and SN\,2007bi. }

\label{fig:OIprofiles}
\end{figure}

\subsection{Optical Spectroscopy ($t > 250$ day)}
\label{sec:late-time}

The defining properties of SN\,2012au are seen in its nebular spectrum
(Figure~\ref{fig:spectra}, top). Emissions from SNe Ib/c at these
epochs are normally optically thin and dominated by [\ion{O}{1}]
\dlambda 6300, 6364, [\ion{Ca}{2}] \dlambda 7291, 7324, and
\ion{Mg}{1}] $\lambda$4571.  In addition to these emissions, however,
SN\,2012au shows persistent P-Cyg absorptions attributable in part to
\ion{Fe}{2} at $\la 2000$~\kms, as well as unusually strong emissions
from \ion{Ca}{2} H\&K, \ion{Na}{1}\,D, and a feature around 7775~\AA\
we identify as \ion{O}{1} $\lambda$7774.

Also seen in the nebular spectrum is a ``plateau'' of emission between
$4000-5600$\,\AA. This emission is due largely to iron-peak elements
(primarily [\ion{Fe}{2}], and some [\ion{Fe}{3}] and [\ion{Co}{2}]),
which is usually only observed with this strength in the late-time
spectra of SNe Ia. Unlike SNe Ia, however, here the [\ion{Fe}{3}] lines are
weak.

In Figure~\ref{fig:OIprofiles} (left), emission line profiles of
SN\,2012au's most prominent late-time spectral features are
shown. There are real differences among the emission line profiles
across elements and their ions. Specific features of note include: (1)
the [\ion{Ca}{2}], \ion{Na}{1}, and \ion{Mg}{1}] lines are not
symmetric about zero velocity and exhibit higher expansion velocities
than [\ion{O}{1}], and (2) the \ion{O}{1} and [\ion{O}{1}] lines
indicate two distinct velocity regions in the ejecta.

The relative strengths of narrow lines from coincident host galaxy
emission in the nebular spectra were measured to estimate the
explosion site metallicity using the method described in
\citet{Sanders12}. From the N2 diagnostic of \citet{PP04}, we meausure
an oxygen abundance of $\log({\rm O/H})+12=8.9$ with uncertainty 0.2
dex.  Adopting a solar metallicity of $\log({\rm O/H})_\odot+12=8.7$
\citep{Asplund05}, the measurement indicates that SN~2012au exploded
in an environment of super-solar metallicity around
$Z\sim1-2~Z_\odot$. This metallicity is significantly higher than any
of the broad-lined SN Ic host galaxies from untargeted surveys
\citep{Sanders12}.

\subsection{Near-Infrared Spectroscopy}

Three epochs of low-resolution, near-infrared spectra spanning 0.82 to
2.51 $\mu$m were obtained using the FoldedPort Infrared Echellette
(FIRE) spectrograph on the 6.5-m Magellan Baade Telescope
\citep{Simcoe08}. The low dispersion prism used in combination with a
$0\farcs 6$ slit yielded a spectral resolution $R \sim 500$ in
$J$-band.  Data were reduced following standard procedures (see, e.g.,
\citealt{Hsiao13}) using a custom-developed IDL pipeline (FIREHOSE).

The reduced near-infrared spectra are plotted in Figure~\ref{fig:spectra}
(bottom). In our $\Delta t_{\rm max} = +31$\,d spectrum, absorptions
around both the \ion{He}{1} $\lambda$1.083$\mu$m and
$\lambda$2.05$\mu$m lines support the identification of \ion{He}{1}
made in the optical spectra. The minima of these absorptions shift
toward longer wavelengths in the later epochs, and the absorption
around 1\,$\mu$m grows in strength as the 2\,$\mu$m absorption
fades. This suggests that the \ion{He}{1} strength diminishes as
other ions possibly including \ion{Si}{1} and \ion{C}{1} gradually
dominate the 1\,$\mu$m absorption as time passes.

Between $\Delta t_{\rm max} = +81$\, and $+318$\,d, the
full-width-at-half-maximum (FWHM) of strong emission features
associated with the \ion{Mg}{1} $\lambda$1.503\,$\mu$m and \ion{O}{1}
$\lambda$1.317\,$\mu$m lines narrow from approximately 5700 \kms\ to
2000 \kms. The presence of these lines and the similarity of their
velocity distribution to the feature around 7775~\AA\ support our
identification of it being associated with \ion{O}{1} $\lambda$7774
(Figure~\ref{fig:OIprofiles}, left).

\begin{figure*}[htp!]
\centering

\includegraphics[width=0.49\linewidth]{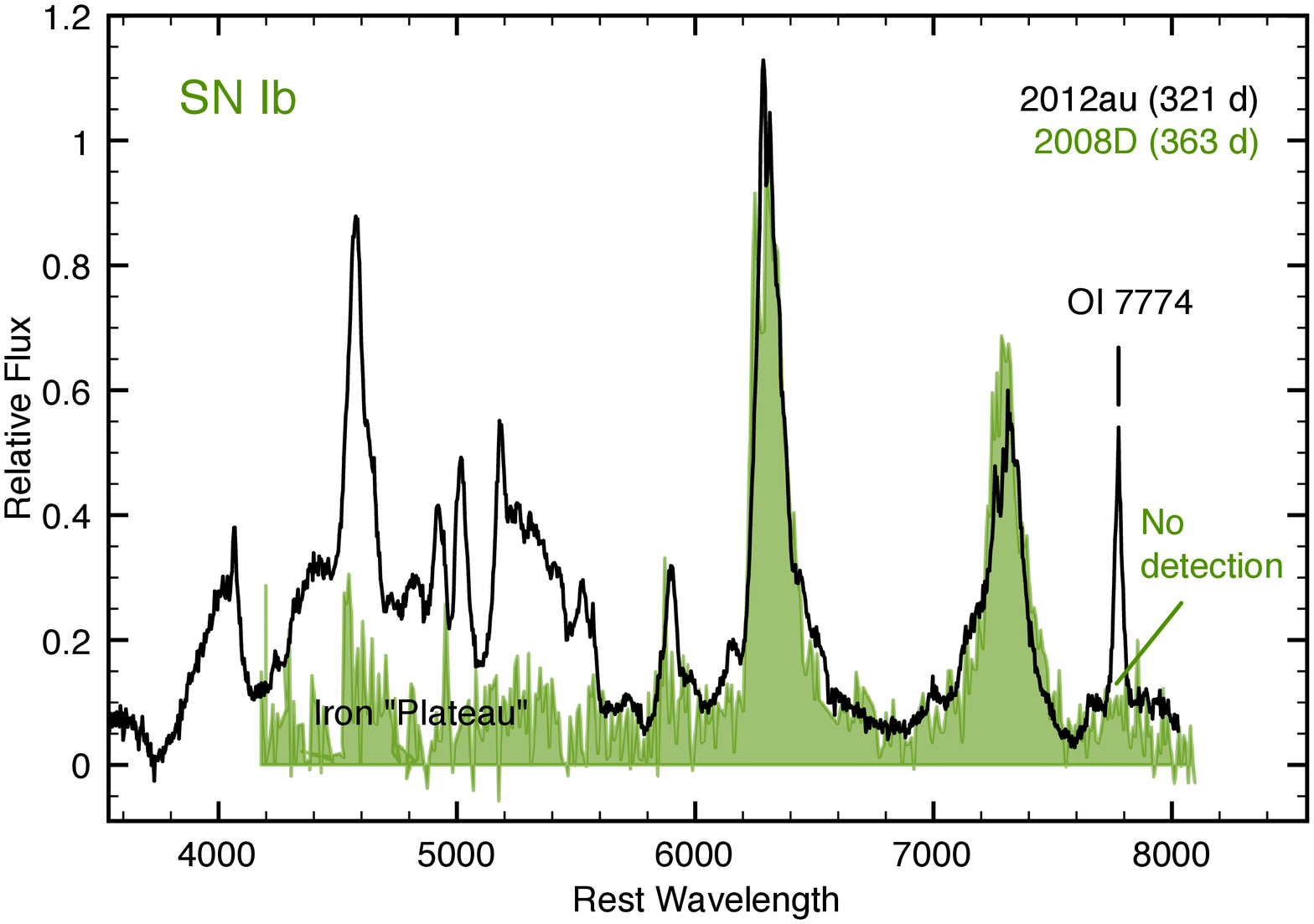}
\includegraphics[width=0.49\linewidth]{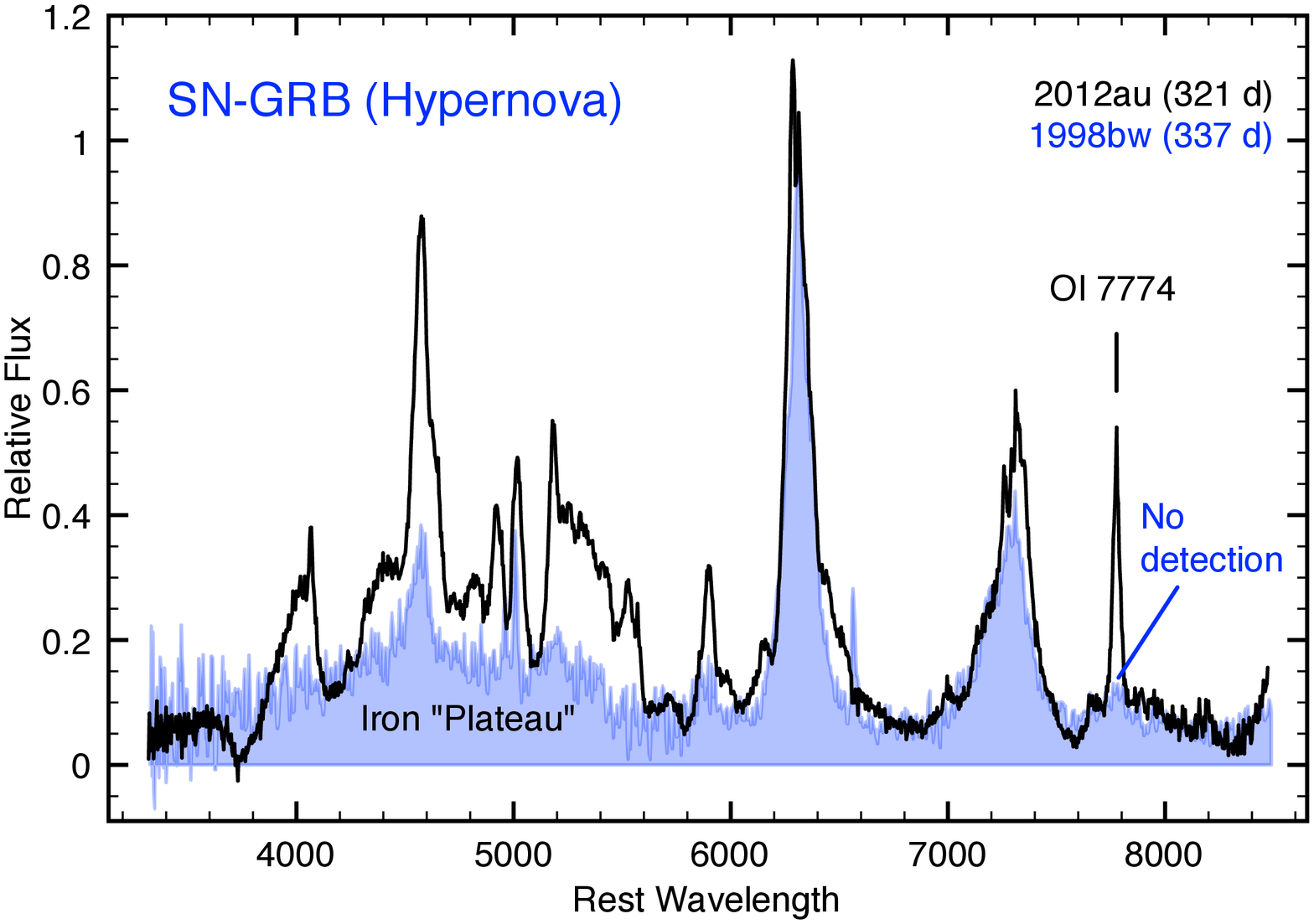}
\includegraphics[width=0.49\linewidth]{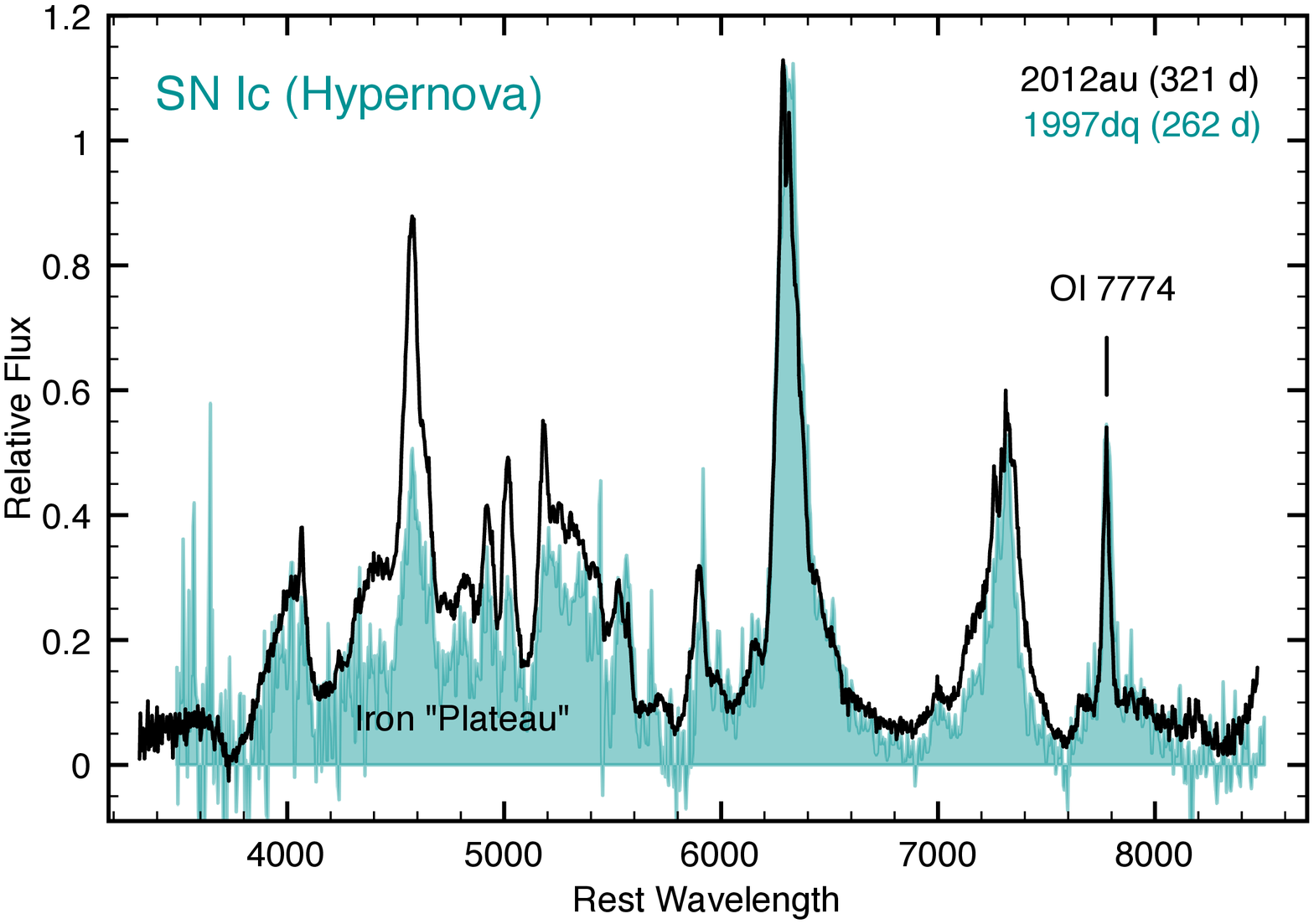}
\includegraphics[width=0.49\linewidth]{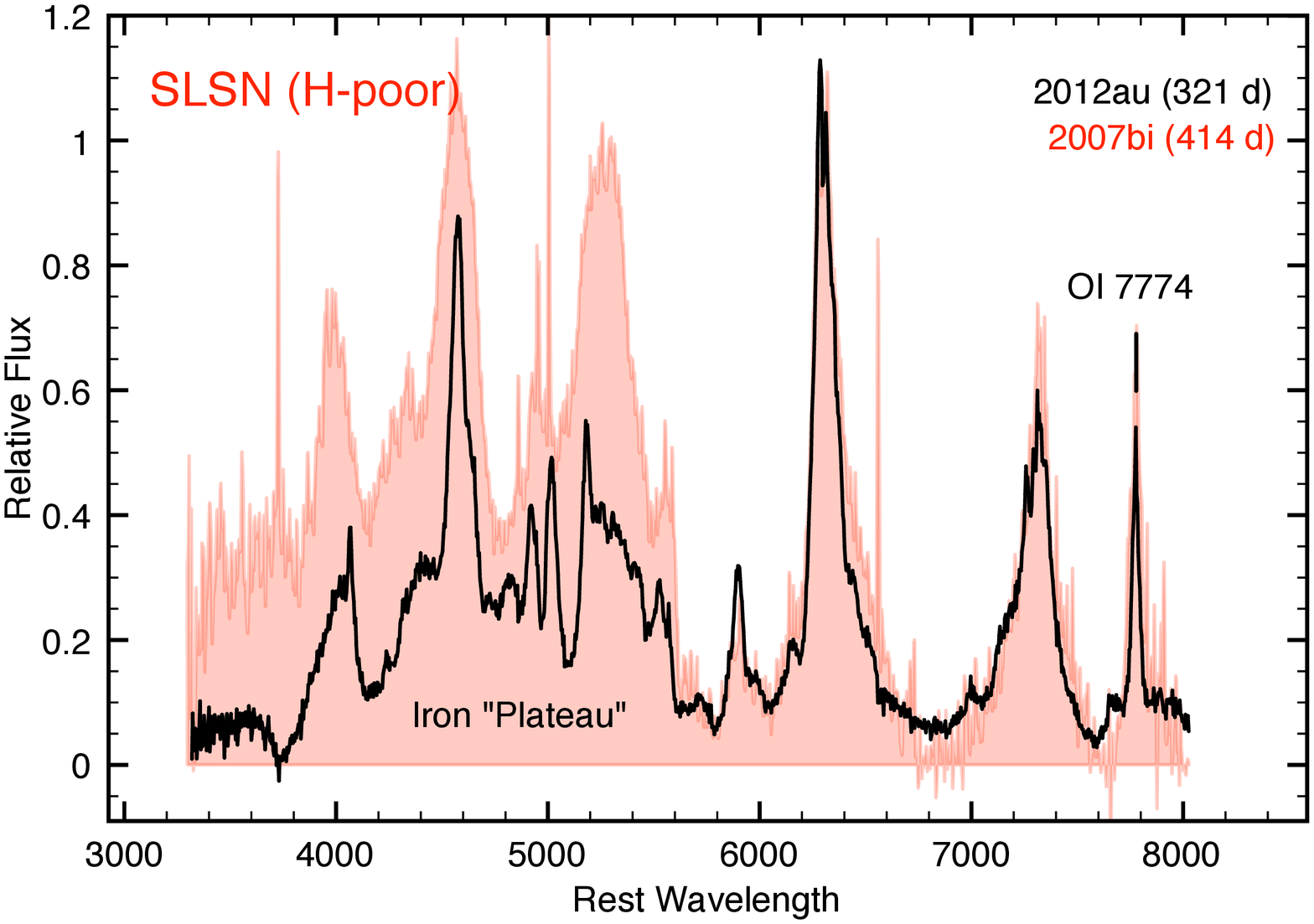}

\caption{Late-time spectrum of SN\,2012au obtained 2013 February 5
  compared to those of other SNe Ib/c. Relative flux has been
  arbitrarily scaled to match the strength of [\ion{O}{1}] \dlambda
  6300, 6364 emission. Top panels show the SN Ib SN\,2008D
  \citep{Tanaka09} and the hypernova SN\,1998bw \citep{Patat01}, where
  spectroscopic evolution is normal and \ion{O}{1} $\lambda$7774 is
  not detected. These SNe are representative of late-time emissions
  from the majority of SNe Ib/c (see, e.g.,
  \citealt{Matheson01,Taubenberger09,Milisavljevic10}).  Bottom panels
  show the hypernova SN\,1997dq \citep{Matheson01} and the SLSN
  SN\,2007bi \citep{Gal-Yam09}, where spectroscopic evolution is slow,
  strong \ion{O}{1} $\lambda$7774 emission is detected, and emission
  from Fe lines forms an emission plateau between $4000-5600$~\AA.}

\label{fig:filledplot}
\end{figure*}

%%%%%%%%%%%%%%%%%%%%%%%%%%%%%%%%%%%%%%%%%%%%%%%%%%%%%%%%%%%%%%%%%%%%%
\section{Discussion}
\label{sec:Discussion}
%%%%%%%%%%%%%%%%%%%%%%%%%%%%%%%%%%%%%%%%%%%%%%%%%%%%%%%%%%%%%%%%%%%%%

\subsection{Extraordinary Late-Time Emission Properties} 

Our UV/optical and near-infrared observations of SN\,2012au show it to
be a slow-evolving energetic supernova with a number of rarely
observed late-time emission properties. SN\,2012au's plateau of iron
emission lines, intermediate-width \ion{O}{1} $\lambda$7774 emission,
persistent P-Cyg absorption features, and prolonged brightness almost
one year after explosion are not observed in the majority of late-time
spectra of SNe Ib/c (Figure~\ref{fig:filledplot}, top).

However, a handful of objects including SN\,1997dq and 2007bi do share
SN\,2012au's rare blend of late-time properties. Aside from some
difference in the strength and velocity widths of the Fe and Mg
emissions below $\approx$$5600$~\AA, above $\approx$$5600$~\AA\
emissions from these SNe are almost indistinguishable
(Figure~\ref{fig:filledplot}, bottom). Like SN\,2012au, these objects
exhibited slow spectroscopic evolution and slowly declining light
curves (see \citealt{Mazzali04}, \citealt{Gal-Yam09}, and
\citealt{Young10}).

\citet{Mazzali04} interpreted the long duration of the photospheric
phase in SN\,1997dq and SN\,1997ef to be the consequence of a sharply
defined two-component ejecta distribution made of an inner,
high-density region, located inside a much lower-density region of
high-velocity ejecta. This was a scenario that had been previously
modeled by \citet{Maeda03}. Mazzali et al.\ concluded that a
significant fraction of Ni-rich material was associated with
velocities below the normal mass cut-off imposed in one-dimensional
explosion models and that this signaled the presence of explosion
asymmetries.

A similar explanation may be applicable to SN\,2012au. The persistent
P-Cyg absorptions and asymmetries between elements and their ions in
the emission line profiles are consistent with expectations of a
moderately aspherical explosion.  Moreover, the absence of
[\ion{Fe}{3}] in the iron plateau region (Figure~\ref{fig:filledplot})
indicates that the density of the Fe-rich region is high and likely
clumped. Thus, it is possible this asphericity was jet-driven, since
models have demonstrated that jetted explosions can significantly
alter density profiles and $^{56}$Ni distributions, especially in the
central region \citep{MaedaNomoto03}.

The presence of the density-sensitive line \ion{O}{1} $\lambda$7774
and its velocity width is especially noteworthy. In the cases of
SN\,2012au and SN\,2007bi, the center of the \ion{O}{1} distribution
sits in the middle of an emission gap of the [\ion{O}{1}] \dlambda
6300, 6364 profile (Figure~\ref{fig:OIprofiles}, right). This, along
with the \ion{Mg}{1} 1.503$\mu$m feature in the near-infrared, is
suggestive of an O- and Mg-rich region with densities that are
collisionally quenching some of the [\ion{O}{1}] \dlambda 6300, 6364
emission.

\subsection{A Unified Explosion Mechanism?}

On the one hand, it seems unlikely that SNe with very different peak
luminosities and light curves can share progenitor systems and
explosion mechanisms (Figure~\ref{fig:lcurves}, right). SN\,2007bi was a
superluminous ($M_B \approx -21$\,mag) explosion that ejected some
$\sim 3-5M_{\odot}$ of radioactive $^{56}$Ni, and was associated with
a 200 $M_{\odot}$ progenitor (ZAMS) located in a low metallicity dwarf
galaxy \citep{Gal-Yam09,Young10}. These properties are very different
from SN\,2012au, which was far less luminous ($M_B \approx -18$\,mag),
produced an order of magnitude less $^{56}$Ni, and was located in a nearby
flocculent spiral we estimate to have super-solar
metallicity. Moreover, we find that a comparison of Geneva stellar
evolution models \citep{Ekstrom12} to results from a preliminary
analysis of pre-explosion \emph{Hubble Space Telescope} images of the
region encompassing SN\,2012au \citep{VanDyk12} places loose limits
on the progenitor star as likely being $<80\;M_{\odot}$.

Despite these differences, however, the combination of observed
optical properties shared between these SNe is still suggestive of a
unified explosion mechanism.  If a connection between these objects
exists, then one of the three currently favored energy mechanisms of
SLSNe powering their slow light curves should extend to
SN\,2012au. One late-time mechanism is interaction with circumstellar
material. Although no narrow emission lines (FWHM $\la 100$ \kms) have
been detected, interaction with H-poor circumstellar shells from
pulsational pair-instability SNe could potentially be applicable
\citep{Chatzopoulos12}. Favoring against this, however, is that
SN\,2012au's radio light curves and SED evolution are most consistent
with blast wave interaction with a steady, homogeneous wind (Kamble et
al., in prep). Another mechanism is injection of energy from a
magnetar, but the presence of strong iron emission, as observed in
SN\,2012au, could be problematic for this scenario \citep{Kasen10}.

A straightforward understanding of SN\,2012au's light curve is
radioactive $^{56}$Ni. The rise and fall of the light curve at
photospheric epochs, and the subsequent slow decline of the light
curve at the rate of $\approx 0.01$ mag d$^{-1}$ up until $\sim
300$\,d, reasonably follow predictions of Ni-Co decay
(Figure~\ref{fig:lcurves}). Possible origins for the above average
$^{56}$Ni production may be either a pair instability explosion or Fe
core collapse of a massive progenitor.  A pair-instability explosion
is unlikely given that the progenitor mass and metallicity we estimate
for SN\,2012au are outside theoretical limits (see
\citealt{Langer07}). Hence, Fe core collapse is the prime
candidate. SN\,2012au's exceptionally high energies suggest that the
explosion may have been aided by magnetohydrodynamic jets brought
about by rapid rotation \citep{Burrows07}.

\section{Conclusions}

We have shown that SN\,2012au is an energetic ($E_{\rm K} \sim
10^{52}$\,erg) explosion having a rarely observed combination of
late-time properties that suggest a link between subsets of energetic
and H-poor SNe and SLSNe. These events, which extend over a large
range of absolute magnitudes ($-21 \la M_B \la -17$ mag), appear to be
observationally connected by a slow spectroscopic evolution expressed
by persistent P-Cyg absorptions, intermediate-width ($\sim 2000$ \kms)
\ion{O}{1} $\lambda$7774 emission approximately a year after
explosion, and slowly declining light curves. We conclude that they
may be unified in a single framework involving the core collapse of a
massive progenitor and a subsequent asymmetric explosion.

%% Future work may look to investigate the presented optical and NIR data
%% with explosion models and nebular emission codes. Such models can
%% explore details of the explosion dynamics and provide estimates of the
%% chemical abundances of the ejecta to more accurately constrain the
%% progenitor structure. In turn, this information could shed additional
%% light on what particular attributes of these rare events separates
%% them from other more commonly observed SNe.

\acknowledgements

We thank an anonymous for informed suggestions that improved the
manuscript, and the Harvard College Observatory for supporting the
Astronomy100 class who were first to classify SN\,2012au. Support was
provided by the David and Lucile Packard Foundation Fellowship for
Science and Engineering awarded to A.M.S. Additional support is from
the NSF under grants AST-0306969, AST-0607438, AST-1008343, and
AST-121196. Observations reported here were obtained at the MMT
Observatory, a joint facility of the Smithsonian Institution and the
University of Arizona, as well as the 6.5 m Magellan Telescopes
located at Las Campanas Observatory, Chile. This paper used the
Weizmann interactive supernova data repository
(\texttt{http://www.weizmann.ac.il/astrophysics/wiserep}).

% \bibliographystyle{yahapj}
% \bibliography{ref_sne}

\clearpage
\newpage

\begin{deluxetable*}{lcccccccl}[htp!]
\footnotesize
\centering
\tablecaption{UV/Optical Photometry}
\tablecolumns{9}
\tablewidth{0pt}
\tablehead{\colhead{MJD} &
           \colhead{$uvm2$}  &
           \colhead{$uvw2$}   &
           \colhead{$uvw1$}   &
           \colhead{$u$}        &
           \colhead{$b$}        &
           \colhead{$v$}        &
           \colhead{SDSS $r'$}        &
           \colhead{Telescope/Instr.}}
\startdata
 56001.74  &     15.91          0.06 &     15.74       0.07  &    14.72       0.06  &    13.52       0.05 &     14.11       0.04  &    13.67       0.04 & \nodata & \emph{Swift}-UVOT \\
 56004.75  &     16.00          0.07 &     15.87       0.07  &    14.87       0.06  &    13.59       0.05 &     14.03       0.04  &    13.56       0.04 & \nodata & \emph{Swift}-UVOT \\
 56005.95  &     15.95          0.09 &     15.91       0.08  &    15.02       0.07  &    13.65       0.05 &     14.02       0.05  &    13.51       0.05 & \nodata & \emph{Swift}-UVOT \\
 56006.80  &     16.09          0.07 &     15.95       0.07  &    15.02       0.06  &    13.78       0.05 &     14.05       0.04  &    13.54       0.04 & \nodata & \emph{Swift}-UVOT \\
 56007.79  &     16.06          0.07 &     15.99       0.07  &    15.13       0.06  &    13.88       0.05 &     14.09       0.04  &    13.51       0.04 & \nodata & \emph{Swift}-UVOT \\
 56011.76  &     16.32          0.07 &     16.24       0.08  &    15.59       0.06  &    14.39       0.05 &     14.38       0.05  &    13.61       0.04 & \nodata & \emph{Swift}-UVOT \\
 56020.18  &     16.57          0.07 &     16.55       0.08  &    16.14       0.06  &    15.46       0.06 &     15.16       0.05  &    14.15       0.04 & \nodata & \emph{Swift}-UVOT \\
 56026.48  &     16.50          0.07 &     16.47       0.08  &    \nodata            &   15.78       0.06 &     15.57       0.05  &    14.61       0.05 & \nodata & \emph{Swift}-UVOT \\
 56029.61  &     16.57          0.07 &     16.56       0.08  &    16.45       0.07  &    15.81       0.06 &     15.66       0.05  &    14.63       0.04 & \nodata & \emph{Swift}-UVOT \\
 56033.62  &     16.59          0.07 &     16.62       0.08  &    16.42       0.07  &    15.95       0.06 &     15.76       0.05  &    14.73       0.04 & \nodata & \emph{Swift}-UVOT \\
 56035.60  &     16.55          0.07 &     16.45       0.08  &    16.32       0.07  &    15.95       0.06 &     15.85       0.05  &    14.84       0.05 & \nodata & \emph{Swift}-UVOT \\
 56038.46  &     16.64          0.07 &     16.56       0.08  &    16.41       0.07  &    15.95       0.06 &     15.86       0.05  &    14.88       0.05 & \nodata & \emph{Swift}-UVOT \\
 56315.43  &     \nodata             &     \nodata           &    \nodata           &    \nodata          &     \nodata           &    \nodata          & 16.95 0.10 & MDM/OSMOS  \\
 56341.52  &     \nodata             &     \nodata           &    \nodata           &    \nodata          &     \nodata           &    \nodata          & 17.94 0.13 & MMT/MMTcam

 \tablecomments{Uncertainties are adjacent to measurements and are at
   the 68\% confidence level. }

\label{tab:photometry}
\end{deluxetable*}

\end{document}